\documentclass[final,twocolumn,12pt]{elsarticle}

\usepackage{amssymb}
\usepackage{comment}
\usepackage{upgreek}
\usepackage{amsmath}
\usepackage{esint}
\usepackage{url}
\usepackage{nicematrix}
\usepackage[hidelinks]{hyperref}
\usepackage{lineno}

\newcommand{\rev}[1]{\textcolor{black}{#1}}

\journal{Optics and Laser Technology}

\begin{document}


\begin{frontmatter}

\title{Simultaneous measurement of refraction and absorption with an  integrated near-infrared Mach-Zehnder interferometer}


\author[telma]{Antonia Torres-Cubillo\corref{atc}}
\ead{atc@ic.uma.es}
\cortext[atc]{Corresponding author}
\author[telma,wmu]{Alejandro Sánchez-Postigo}
\author[uit]{Jana Jágerská}
\author[telma,ibima]{J. Gonzalo Wangüemert-Pérez}
\author[telma,ibima]{Robert Halir}


\affiliation[telma]{organization={Telecommunication Research Institute (TELMA), University of Malaga},
            addressline={Louis Pasteur 35}, 
            city={29010 Malaga},
            country={Spain}
}

\affiliation[wmu]{organization={Currently with the Department for Quantum Technology, University of Münster},
            addressline={Heisenbergstra{\ss}e 11}, 
            city={48149 Münster},
            country={Germany}
}

\affiliation[uit]{organization={Department of Physics and Technology, UiT The Artic University of Norway},
            city={9037 Tromsø},
            country={Norway}
}

\affiliation[ibima]{organization={IBIMA-BIONAND},
            addressline={Malaga Tech Park}, 
            city={29010 Malaga},
            country={Spain}
}


\begin{abstract}

Most integrated evanescent-field photonic sensors measure changes in either the real part or the imaginary part of the complex refractive index of the sample, i.e., refraction or absorption. Here we propose and experimentally demonstrate a near-infrared sensor based on a silicon nitride Mach-Zehnder interferometer which provides a direct measurement of the complex refractive index. Our architecture exhibits a high sensitivity, achieving limits of detection below $2\cdot 10^{-6}\,\mathrm{RIU}$ for both the real and imaginary parts of the refractive index. We furthermore show that our sensor can be employed as an integrated dispersion spectrometer. 

\end{abstract}

\begin{keyword}
Integrated photonics \sep Refraction sensor \sep Absorption sensor \sep Dispersion spectroscopy \sep Mach-Zehnder interferometer \sep Coherent phase detection
\end{keyword}

\end{frontmatter}


\section{Introduction} \label{sec:intro}

In the last decade, photonic integrated sensors have experienced a significant growth as alternative methods for ultra-sensitive, quantitative and label-free detection of analytes such as biological molecules \cite{Dhote2022} or \rev{gases} \cite{Jingwen2020,Seba2021}. Sensors based on CMOS-compatible platforms can furthermore benefit from large-scale production leveraging the infrastructure of the mature semiconductor industry, thus facilitating the development of cost-effective point-of-care devices and lab-on-chip platforms \cite{Hinkov2022}. The detection of the target analyte is performed by correlating its concentration with changes either in the real ($n$) or the imaginary ($k$) part of the complex refractive index of the sample. The former is associated with phase changes of the lightwave, while the latter is related to light absorption following the Beer-Lambert law.

Among real-part refractive index sensors, the two most common architectures are resonators and interferometers. Resonant architectures such as rings \cite{deGoede21,Sakib2022} and discs \cite{Aghaei2023} have a strong potential for miniaturization and multiplexation \cite{Holler2023}, but, as they operate by tracking changes in the resonance wavelength, they usually require tunable laser sources or spectral analysers, which could elevate the complexity and cost of the overall system. Interferometric architectures measure variations in the refractive index by detecting changes in the phase shift between a sensing and a reference signal and can therefore be operated at a fixed wavelength. Mach-Zehnder interferometers (MZI) are one of the commonly used configurations, and they have been extensively exploited for a precise and sensitive detection of both inert and biological analytes \cite{Dullo15,Besselink2022}. Furthermore, coherently-detected MZIs, aided by optical combiners, provide a linear and unambiguous readout, thus avoiding problems associated with the sinusoidal nature of the interferometric signal, such as sensitivity fading and output ambiguity, without needing heaters or external modulation schemes \cite{Halir2013,Leuermann20,SANCHEZRAMIREZ2024109813}. Recently, bimodal interferometers have emerged as an interesting alternative. Instead of using two different physical paths, they are based on just one waveguide which supports two orthogonal modes. With this approach they are able to provide results comparable to those obtained by MZIs while substantially reducing the sensor footprint \cite{Torrijos2022,MiBimodal}.

Absorption sensors are widely employed for both near (NIR) and mid-infrared (MIR) sensing applications \cite{Huck2020}. Although they can operate at a fixed wavelength to detect dynamic absorption changes, many rely on the analysis of the absorption spectrum of a sample to achieve specificity. Integrated spectroscopic sensors in the NIR benefit from mature manufacturing technologies and the availability of cost-effective sources and detectors. However, due to the weak absorption of NIR spectral features, which are often overlapped overtones of different chemical species, they usually require performing a multivariate data analysis to obtain the target spectral response \cite{BorisNIR2014,VanDamme2023}. In strongly absorbing regions such as the MIR, where many molecules present fingerprint spectra, the detection of specific analytes can be more straightforward. As a consequence, sensors based on spiral or straight highly-sensitive waveguides achieve ultra-low \rev{limits of detection} without adding further architectural complexity \cite{Vinita2019,Marek2021,Andreacaffeine}. In order to push overall sensing capabilities, some more sophisticated configurations, such as multimode interferometers (MMI) \cite{Bodiou2020} or frequency combs \cite{Pan2023}, are emerging. 

\begin{figure*}[htb]
\centering
\includegraphics[width=0.85\linewidth]{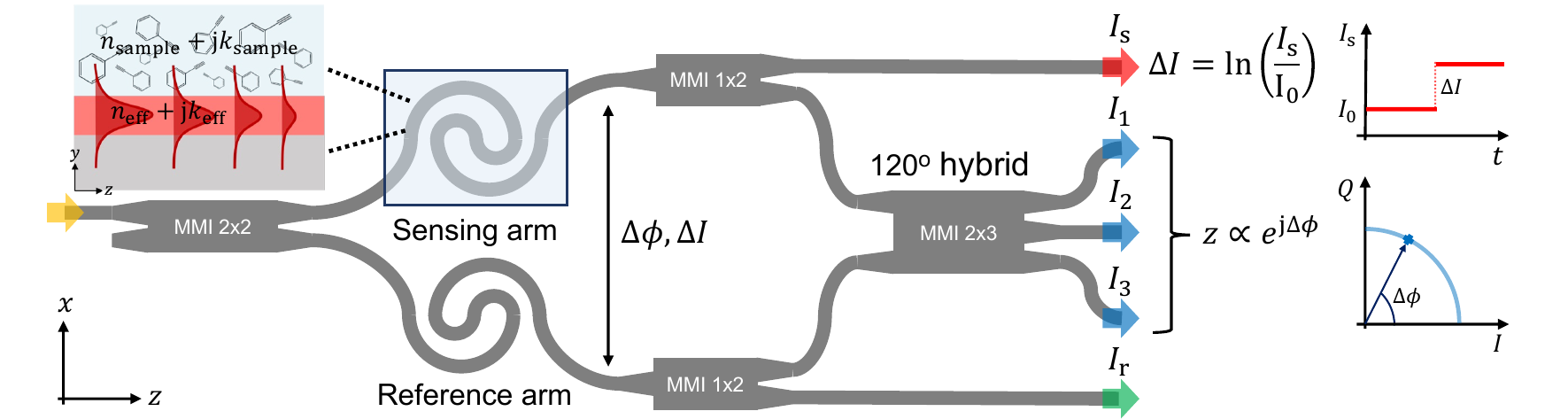}
\caption{Schematic representation of the MZI-based sensor employed in this work. Input light is split into a sensing and a reference path. Due to the overlap between the evanescent field of the sensing mode and the analyte, \rev{the sensing signal accumulates an attenuation ($\Delta I$) relative to the background signal and a phase shift ($\Delta \phi$) relative to the reference signal after propagating through the MZI arms}. Recombination into an optical hybrid enables the coherent detection of this phase shift. Direct intensity measurements are used to determine the absorption.}
\label{fig:diagram}
\end{figure*}

Refraction and absorption sensors provide partial information about the sample, as they only record one component of its complex refractive index. This may be sufficient for many applications, specially for those where one the components of the complex refractive index of the sample is easier to measure than the other, as is the case for the refraction of most liquids in the NIR and the absorption of gasses in the MIR. However, simultaneous detection of the real and the imaginary parts of the index in a single measurement would add a further insight into the nature of the substance, which may be interesting in some scenarios, such as the analysis and detection of volatile organic compounds (VOC) like tert-butyl methyl ether, which show comparably strong peaks in their dispersion and absorption spectra \cite{NIST}, and could benefit from their combined measurement. Furthermore, such complex refractive index sensors would exhibit an improved robustness by introducing redundancy, and provide versatility in a unified architecture, as they combine the capabilities of refraction and absorption sensors. Even though Kramers-Kronig relationships can be employed to obtain one part of the complex refractive index from the other \cite{Kop1997}, this requires measurements over a very wide spectral range \cite{Mayerhofer2020}. Despite their scientific interest, examples of sensors which simultaneously detect refraction and absorption are still scarce in literature: a dual-comb-based free space implementation was reported in \cite{Lisak2022}, and, to the best of our knowledge, the only integrated solutions presented so far are based on Young interferometers in the visible range (VIS) \cite{Zhou18}, ring resonators \cite{Briano20} in the MIR and a photonic crystal microdisk in the VIS for multiplexed detection \cite{Blasi23}. However, these approaches require a relatively complex readout with a camera or a tunable wavelength source. More importantly, the limit of detection they achieve is limited to the order of $10^{-5}\,\mathrm{RIU}$ in either the real or imaginary part of the refractive index.  

Here, we propose and experimentally demonstrate a MZI sensor capable of simultaneously measuring refraction and absorption changes around a central wavelength of $1.55\,\mathrm{\upmu m}$, fabricated in a commercial silicon nitride foundry [see Fig.\,\ref{fig:diagram}]. Our architecture enables coherent phase retrieval and direct measurement of absorption with simple photodiodes and a fixed wavelength laser, achieving limits of detection of the order of $10^{-6}\,\mathrm{RIU}$ for both the real and imaginary part of the refractive index. To the best of our knowledge, this is the best performance for integrated complex refractive index sensing yet reported. Furthermore, we show the potential of our approach for spectroscopic dispersion measurements.

This work is organized as follows. In section \ref{sec:funds}, we explain the fundamentals of evanescent-field interferometric photonic integrated sensors, particularized for our architecture. Our sensing setup and the sensing experiments are detailed in section \ref{sec:exp}, along with an analysis of results. Conclusions are drawn in section \ref{sec:concl}. Additionally, in \ref{ap:fringes}, we included a model to explain and mitigate the presence of fringes in our measurements. 

\section{Fundamentals of complex refractive index sensing} \label{sec:funds}

In the MZI configuration schematized in Fig.\,\ref{fig:diagram}, input light is split into a sensing and a reference arm. The sensing arm is exposed to the sample while the reference arm is isolated from it. In the sensing arm, the evanescent field of the mode overlaps with the sample, thus making the mode sensitive to changes in the composition of the analyzed substance, as illustrated in the inset of Fig.\,\ref{fig:diagram}. Consequently, changes in the complex refractive index of the sample ($\Delta n_\mathrm{sample}+\mathrm{j}\Delta k_\mathrm{sample}$) will induce proportional changes in the complex effective index of the mode, $\Delta n_\mathrm{eff}+\mathrm{j} \Delta k_\mathrm{eff}=\Gamma (\Delta n_\mathrm{sample}+\mathrm{j}\Delta k_\mathrm{sample})$. The proportionality constant is given by the waveguide confinement factor $(\Gamma)$  \cite{Gonzalo2019}, which is equal for the real and the imaginary parts of the complex refractive index, i.e. $\Gamma \sim \partial n_\mathrm{eff}/\partial n_\mathrm{sample} = \partial k_\mathrm{eff}/\partial k_\mathrm{sample}$ \cite{Robinson08}.

A change in the concentration ($c$) of the analyte in the sample will induce a wavelength-dependent change in the complex refractive index of the sample: $\Delta n_\mathrm{sample}(\lambda,c)+\mathrm{j}\Delta k_\mathrm{sample}(\lambda,c)$. After propagating through the arms of the MZI, with an equal length $L$, the sensing and reference modes accumulate a difference both in phase and attenuation. The relative phase shift is given by
\begin{equation}
    \Delta\phi(\lambda,c)=\frac{2\pi}{\lambda}L\left(n_\mathrm{eff}(\lambda,c)-N(\lambda)\right)
    \label{eq:DeltaPhi}
\end{equation}
where $n_\mathrm{eff}(\lambda,c)=n_\mathrm{eff}(\lambda,0)+\Gamma \Delta n_\mathrm{sample}(\lambda,c)$ is the effective index of the sensing waveguide and $N$ is the effective index of the reference waveguide. With fixed wavelength $(\lambda_0)$ operation we find that the differential phase shift with and without the analyte is:
\begin{equation}
 \Delta\phi(\lambda_0,c)-\Delta\phi(\lambda_0,0)  = \frac{2\pi}{\lambda}L\Gamma \Delta n_\mathrm{sample}(\lambda_0,c), 
\end{equation}
thus providing a direct measurement of $\Delta n_\mathrm{sample}$, which we can correlate with the concentration, $c$. To unambiguously obtain
$\Delta\phi$, both arms of the interferometer are combined in a $2\times3$ MMI. A complex signal $z\propto\exp(\mathrm{j}\Delta\phi)$ in the Inphase-Quadrature (IQ) plane [see Fig.\,\ref{fig:diagram}] is generated by multiplying the output intensities $(I_1,I_2,I_3)$ of the MMI with a coefficient matrix \cite{Halir2013,Reyes-Iglesias12}. This readout technique yields a constant sensitivity and is robust against noise, because the measurement of the phase is orthogonal to the radial noise component \cite{Molina19}. Moreover, simple calibration algorithms can be employed to correct deterministic hardware errors \cite{Halir2013,Reyes-Iglesias12}. While it is theoretically possible to obtain absorption changes from the amplitude of the IQ signal, this is numerically challenging, as this magnitude is highly sensitive to non-linear distortion and cross-talk. A direct assessment of absorption is thus preferred.

We enabled absorption measurements by introducing a splitter based on a $1\times2$ MMI at the output of each arm of the interferometer [see Fig. \ref{fig:diagram}]. We can define the intensity change with and without the analyte as
\begin{multline}
    \Delta I (\lambda,c) = \ln\left(\frac{I_\mathrm{s}(\lambda,c)}{I_\mathrm{s}(\lambda,0)}\right) = \\-\frac{4\pi}{\lambda}L\Gamma \Delta k_\mathrm{sample}(\lambda,c),
    \label{eq:DeltaI}
\end{multline}
where $I_\mathrm{s}$ is the output intensity of the sensing arm and $\Delta k_\mathrm{sample}(\lambda,c)=k_\mathrm{sample}(\lambda,c)-k_\mathrm{sample}(\lambda,0)$. The reference output intensity $I_\mathrm{r}$ can be employed as a referential measurement which may help to correct drifts like those caused by temperature or coupling fluctuations. Note that the sensitivity to imaginary refractive index changes $S_\mathrm{k}=|\partial I/\partial k_\mathrm{sample}| = 2(2\pi/\lambda)L\Gamma$ is twice the value of $S_\mathrm{n}$. 

\begin{figure*}[htb]
\centering
\includegraphics[width=0.7\linewidth]{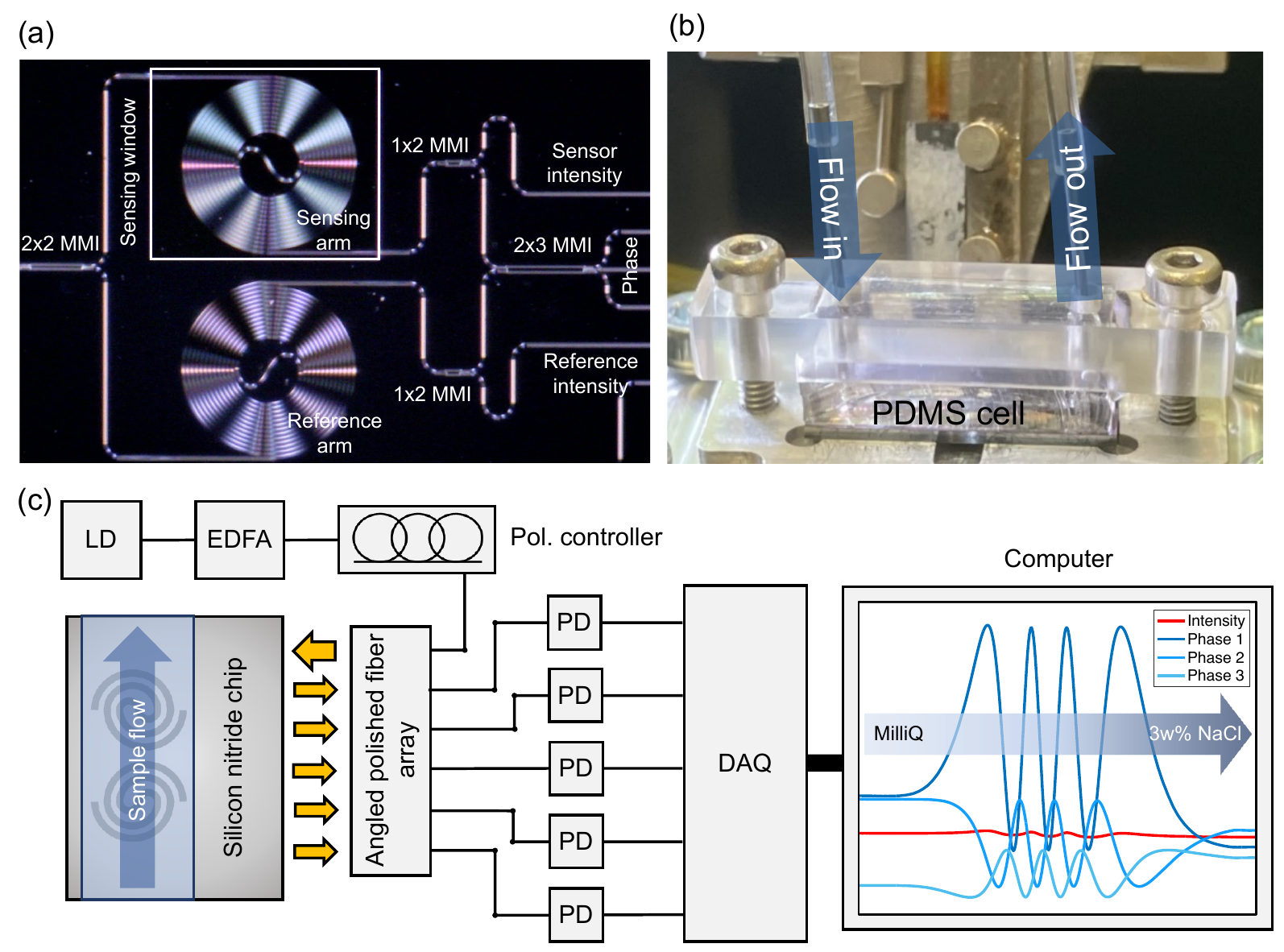}
\caption{(a) Microscope picture of one of the sensors used in this work. Its main components are spiral waveguides and MMIs for light splitting and recombination. (b) Front view of the microfluidic flow-cell placed over the sensing area of the chip. A withdrawal pump is employed to flow the samples through the channel. (c) Schematic representation of the measurement setup. The input signal is generated by a laser (LD), amplified by an EDFA and carried through a polarization controller into a fiber array. Coupling to and from the chip is performed via grating couplers. The five output signals are photodetected (PD), digitalized (DAQ) and processed with Matlab. A fragment of the detected signals during a bulk sensing experiment is shown.}
\label{fig:setup}
\end{figure*}

Although the possibility of single-wavelength operation is a great advantage of our architecture, we can also scan the wavelength over a certain bandwidth and employ our sensor as a spectrometer. By analyzing the spectral responses described in Eqs.\,\eqref{eq:DeltaPhi}-\eqref{eq:DeltaI}, simultaneous dispersion and absorption spectroscopy is possible.

\section{Experimental results} \label{sec:exp}

We designed our sensors for the commercial silicon nitride platform provided by Cornerstone \cite{Cornerstone}, with a 300 nm silicon nitride device layer, 3 $\upmu\mathrm{m}$ buried oxide layer (BOX) and a 2 $\mathrm{\upmu m}$ silicon dioxide cladding. All our structures are manufactured via deep-UV projection lithography in a single step of reactive ion etching. The top cladding is removed from the sensing areas to enable light-sample interaction. The central wavelength is $1.55\,\upmu\mathrm{m}$ and the employed polarization is TE. A microscope picture of a sensor is shown in Fig.\,\ref{fig:setup}(a). The sensing and reference waveguides are $1\,\mathrm{\upmu m}$ wide and equal in length. Sensors with two different interaction lengths (6.72 mm and 16.14 mm) were fabricated. To achieve a compact layout we spiralled the waveguides following the pattern proposed in \cite{Simard13}, with a minimum bend radius of $36.2\,\mathrm{\upmu m}$ and a separation of $15\,\mathrm{\upmu m}$ between adjacent waveguides. A theoretical confinement factor $\Gamma=0.22$ for water-based samples was calculated via FemSIM (RSoft, Synopsis \cite{RSoft}) simulations. \rev{The effect of temperature fluctuations in the complex effective index was quantified, resulting in the thermo-optical sensitivities of $\partial n_\mathrm{eff}/\partial T-\partial N/\partial T=-2.35\cdot10^{-5}\,\mathrm{RIU/^\circ C}$ for the real part and $\partial k_\mathrm{eff}/\partial T=-1.11\cdot10^{-7}\,\mathrm{RIU/^\circ C}$ for the imaginary part, the latter being dominated by the effect of water absorption.} Light is split by $2\times2$ \rev{($8.8\,\mathrm{\upmu m}$ wide, $172\,\mathrm{\upmu m}$ long)} and $1\times2$ \rev{($8.8\,\mathrm{\upmu m}$ wide, $43\,\mathrm{\upmu m}$ long)} MMIs, recombined by a $2\times3$ \rev{($13.3\,\mathrm{\upmu m}$ wide, $256\,\mathrm{\upmu m}$ long)} MMI and coupled to/from the chip via surface grating couplers. A custom PDMS microfluidic flow-cell, connected to a withdrawal pump, and a methacrylate holder [see Fig.\,\ref{fig:setup}(b)] were designed to make the sample flow over the sensor, achieving both full coverage of the sensing areas and hermetical sealing. 

The optics and electronics setup is schematized in Fig.\,\ref{fig:setup}(c). Input light is generated by a laser source (Santec TSL-770), amplified by an erbium-doped fiber amplifier (EDFA) (IPG Photonics EAD-500-C) and coupled into the chip with an angled-polished fiber array (O/E Land). A manual polarization controller (Thorlabs FPC562) is employed to optimize polarization. The five output signals of the sensor are out-coupled to the fiber array, detected by five independent amplified photodetectors (Thorlabs PDA10CS2) and digitalized by a 24-bit DAQ module (NI 9239). Signal processing tasks, including filtering, calibration and readout, are performed by in-house Matlab scripts.

\subsection{Fixed wavelength measurements}

\begin{table}
\begin{center}
\caption{Calculated real and imaginary refractive index changes for the NaCl solutions employed in this work. Pure water is used as a reference.}
\label{tab:nk}
\begin{NiceTabular}{ c c c }
\\
\hline
c [w\%] & $\Delta n_\mathrm{sample}\,\mathrm{[RIU]}$ & $\Delta k_\mathrm{sample}\,\mathrm{[RIU]^\dagger}$ \\
\hline
1.5 & $2.44\cdot10^{-3}$ & $-8.33\cdot10^{-6}$ \\
3 & $4.92\cdot10^{-3}$ & $-1.67\cdot10^{-5}$ \\
6 & $9.97\cdot10^{-3}$ & $-3.33\cdot10^{-5}$ \\
12 & $2.04\cdot10^{-2}$ & $-6.67\cdot10^{-5}$ \\
\hline
\end{NiceTabular}
\end{center}
$^\dagger$ Estimation from \cite{Li2015}.
\end{table}

\begin{figure}[]
\centering 
\includegraphics[width=\linewidth]{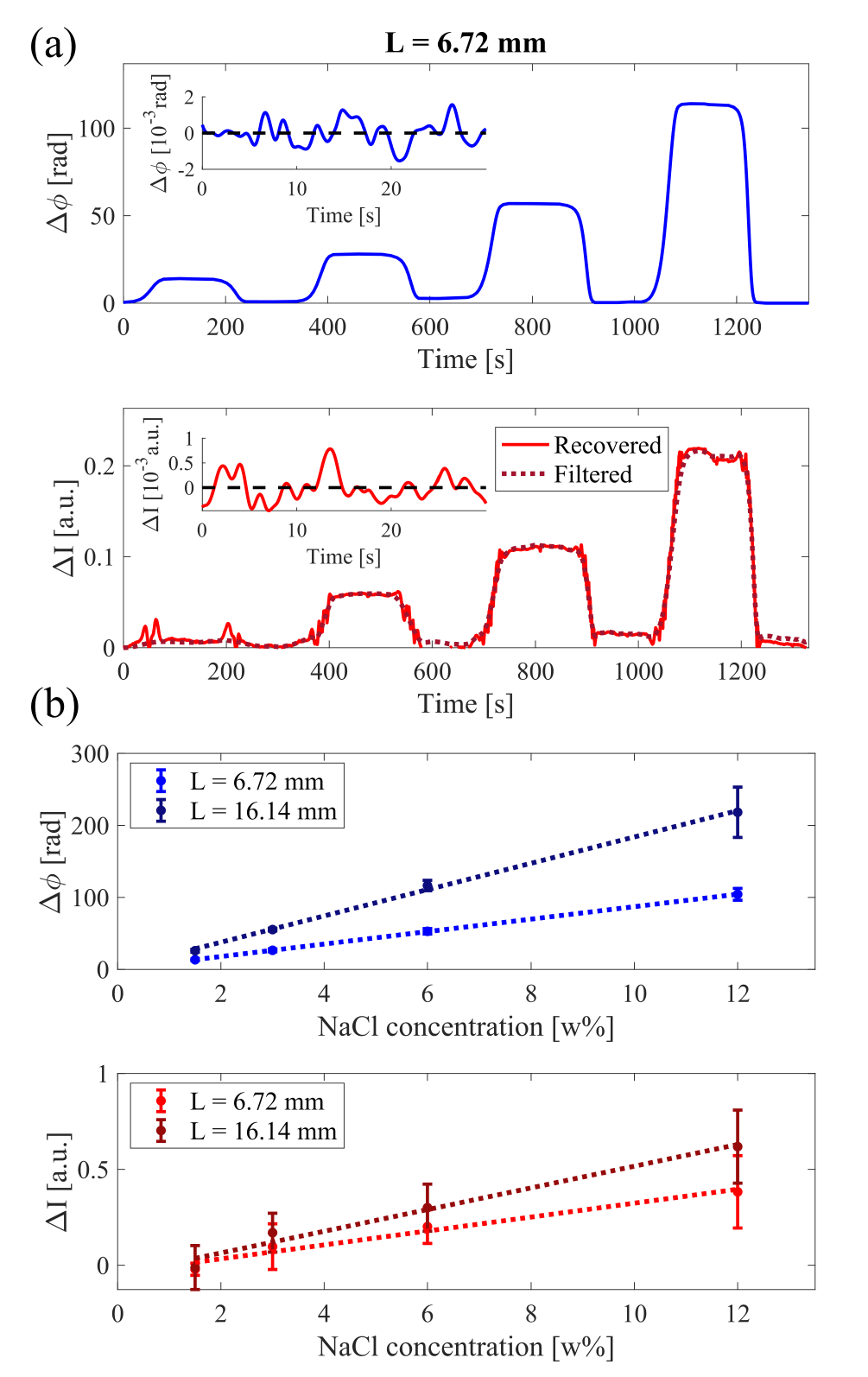}
\caption{\rev{(a)} Phase (upper) and intensity (lower) readout of a 6.7 mm sensor during a bulk sensing experiment at $\lambda_0=1.55\,\mathrm{\upmu m}$. Spurious oscillations on the recovered intensity signal can be filtered with minimum phase algorithms. (b) Averaged phase and intensity sensor responses for two different lengths, with their lineal fit (dotted line). Phase shift error bars for small concentrations are too short to be easily noticed.}
\label{fig:readout}
\end{figure}

Bulk sensing experiments at a fixed wavelength ($\lambda_0=1.55\,\mathrm{\upmu m}$) were performed for the two different sensing lengths and replicated in three chips. We prepared four solutions of sodium chloride (NaCl) diluted in purified water (MilliQ) with weight-percentage (w\%) concentrations of 1.5w\%, 3w\%, 6w\% and 12w\%, whose real refractive indexes were calculated following the model in \cite{Saunders16}. Unfortunately, due to the strong water absorption in this wavelength region, fully accurate absorption data of such NaCl solutions is not widely available in the literature, as the differential change in absorption induced by NaCl is comparatively small. We estimated the imaginary refractive index change generated by our samples based on the data provided in \cite{Li2015}. The real and imaginary index change values assumed in this work are shown in Table \ref{tab:nk}, where the refractive index of pure water at the design wavelength is $1.3154+\mathrm{j}1.49\cdot10^{-4}\,\mathrm{RIU}$ \cite{Kedenburg12}. Here it should be observed that the changes in the real refractive index are around two orders of magnitude larger than in their imaginary counterpart. 

Each sample was flowed over the sensor at a constant flow rate of $30\,\mathrm{\upmu L/min}$, separated by a pure MilliQ buffer. \rev{As the theoretical thermal sensitivity of the sensor is low, temperature control was not implemented in this set of experiments.} The upper panel in Fig. \ref{fig:readout}(a) shows the phase response of the $6.7\,\mathrm{mm}$ long sensor during one of these experiments. A blind calibration of the IQ signal based on geometrical properties \cite{Halir2013} was used to correct for hardware errors, followed by averaging with a $2\,\mathrm{s}$ integration time, yielding a very clean signal. 

The unprocessed intensity signal exhibited oscillations in the time intervals where the gradual transition between buffer and sample was taking place, i.e. during the gradual change in analyte concentration. We attributed these fringes mostly to an undesired cross-talk between the adjacent output signals $I_\mathrm{s}$ and $I_\mathrm{1}$ when coupled into the fiber array [see Fig.\,\ref{fig:diagram}]. We developed a simplified cross-talk model, explained in \ref{ap:fringes}, which enabled us to recover the intensity signal with significantly dampened oscillations. The resulting signal is shown in the lower panel of Fig.\ref{fig:readout}(a) (solid line) for the same $6.7\,\mathrm{mm}$ long sensor. Part of the remaining oscillations in the intensity may be caused by spurious reflections, which would create a Fabry-Pérot-like cavity with a changing effective length caused by the variation of the real part of the effective refractive index. It is possible to further reconstruct the sensor response by applying a minimum phase algorithm to the recovered intensity, thus computing a time-domain reflectogram and filtering the unwanted echoes \cite{halir2009}. The improvement in signal quality is shown in lower panel of Fig.\,\ref{fig:readout}(a) (dotted line). Nevertheless, the quantitative results presented in this work are extracted from the unfiltered intensity signal, as this provided a direct comparison between the phase and intensity readouts.

\begin{table*}
\begin{center}
\caption{Average bulk sensing performance of the \rev{two} measured sensors \rev{in three different chips}.}
\label{tab:results}
\begin{NiceTabular}{ c c c c c c c}
\hline
Length [mm] & $S_\mathrm{n}\,\mathrm{[rad/RIU]}$ & $S_\mathrm{k}\,\mathrm{[a.u./RIU]}$ & $\sigma_\mathrm{n}\,\mathrm{[rad]}$ & $\sigma_\mathrm{k}\,\mathrm{[a.u.]}$ & $\mathrm{LOD_n [RIU]}$ & $\mathrm{LOD_k [RIU]}$ \\
\hline
6.72 & $5.1\cdot10^{3}$ & $6.5\cdot10^{3}$ & 0.0032 & 0.0027 & $1.9\cdot10^{-6}$ & $1.2\cdot10^{-6}$ \\
16.14 & $11\cdot10^{3}$ & $ 10\cdot10^{3}$ & 0.0082 & 0.0075 & $2.3\cdot10^{-6}$ & $2.2\cdot10^{-6}$ \\
\hline
\end{NiceTabular}
\end{center}
\end{table*}

Figure \ref{fig:readout}(b) summarizes the sensitivity results obtained after measuring the 6.7 mm and the 16.14 mm sensors in three different chips. We considered the baseline intensity $I_\mathrm{s}(\lambda_0,0)$ [see Eq.\,\eqref{eq:DeltaI}] of each individual pulse as the one of the immediately preceding MilliQ segment. The sensors are very linear, with $R^2>0.9979$ and $R^2>0.9938$ for phase and intensity readout respectively. To calculate the sensitivity to real and imaginary refractive index changes, we fitted the phase and intensity saturation values to the calculated complex refractive index changes induced by the samples [see Tab.\,\ref{tab:nk}]. Averaged results are shown in Tab.\,\ref{tab:results}. The obtained experimental values for sensitivity to the real part of the refractive index, $S_\mathrm{n}$, are in close agreement with the theoretical $S_\mathrm{n,theo}=5.5\cdot10^3\,\mathrm{rad/RIU}$ and $S_\mathrm{n,theo}=14\cdot10^3\,\mathrm{rad/RIU}$ expected for the 6.7 mm and 16.14 mm sensors respectively. However, the measured values for the sensitivity to the imaginary part of the refractive index are smaller than the theoretically  predicted values $S_{k\mathrm{,theo}}=2S_\mathrm{n,theo}$. This is probably because we slightly overestimated the changes in the imaginary refractive index given in Tab.\,\ref{tab:nk}.

 The limit of detection (LOD) was calculated following the $3\sigma$ criteria. For that matter, we evaluated the noise over 30-second-long fragments of the steady-state signals [see insets in Fig.\,\ref{fig:readout}(a)], resulting in the \rev{average} values shown in Tab.\,\ref{tab:results}. \rev{The strongest contributions to the noise floor are shot noise from the detector, thermal noise from the electrical components, quantization noise from the digitalization stage and residual mechanical noise \cite{Jonas19LOD}.} Regarding the evolution of performance metrics with the interaction length, although sensitivity increases for the longest sensor, as was expected, this is not yielding a better LOD. This is caused by a lower signal-to-noise ratio due to higher losses in the sensing waveguide. Specifically, a differential loss around $11.9\,\mathrm{dB/cm}$ between the sensing and reference waveguide was calculated from the contrast of the interferometric signals as in \cite{Sarmiento2016}. This value can be attributed to water absorption, which causes $11.6\,\mathrm{dB/cm}$ losses. In fact, if we approximate the theoretical optimum length as the inverse of losses, i.e. $L_\mathrm{opt}\sim1/(\Gamma\alpha)$ \cite{Molina19}, where $\alpha$ is the absorption coefficient of water, we get a value of approximately $4\,\mathrm{mm}$, which implies that compact sensors would get the best results in this scenario. Indeed, compared to other results on complex refractive index sensing reported in the literature, our  shortest sensor achieves the best combination of limits of detection, as shown in Tab.\,\ref{tab:comp}.

\begin{table}
\begin{center}
\caption{LODs of integrated complex refractive index sensors operating in different wavelength regimes.}
\label{tab:comp}
\begin{NiceTabular}{c c c}
\hline
Reference & $\mathrm{LOD_n [RIU]}$ & $\mathrm{LOD_k [RIU]}$ \\
\hline
\cite{Zhou18} (VIS) & $1.7\cdot10^{-5}$ & $1.6\cdot10^{-6}$ \\
\cite{Briano20} (MIR) & $8\cdot10^{-6}$ & $1.3\cdot10^{-5}\,^\dagger$\\
\cite{Blasi23} (VIS) & $3\cdot10^{-5}$ & $2\cdot10^{-4}$ \\
This work (NIR) & $1.9\cdot10^{-6}$ & $1.2\cdot10^{-6}$ \\
\hline
\end{NiceTabular}
\end{center}
$^\dagger$ Corresponds to 0.1\% $\mathrm{CO_2}$ at $4234.7\,\mathrm{nm}$ \cite{HITRAN}.
\end{table}

\subsection{Spectroscopic measurements}

\begin{figure}[htb]
\centering 
\includegraphics[width=\linewidth]{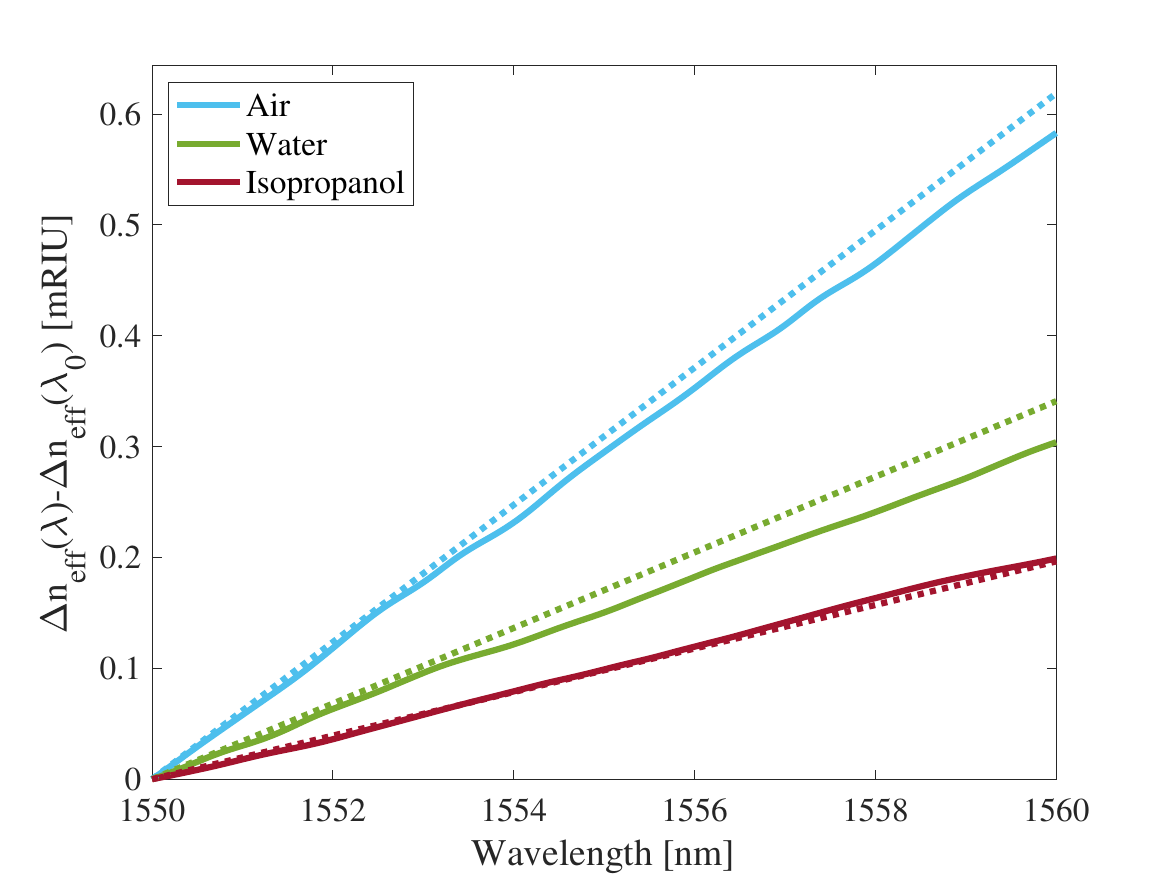}
\caption{Effective index difference as a function of the wavelength for different cladding substances. Simulated values assuming dispersive waveguide materials and samples are plotted in dotted lines.}
\label{fig:sweep}
\end{figure}

We furthermore explored the capabilities of our sensors for dispersion spectroscopy. We scanned the wavelength between $1.55\,\mathrm{\upmu m}$ and $ 1.56\,\mathrm{\upmu m}$, using three different substances as sensor cladding: air, MilliQ water and isopropanol. The change in the real effective index of the mode, $\Delta n_\mathrm{eff}(\lambda) = n_\mathrm{eff}(\lambda,c)-N(\lambda)$, was extracted from the phase measurements using  Eq.\,\eqref{eq:DeltaPhi}. We can correlate this measurement to the dispersion spectrum of the sample by referring it to the starting wavelength, i.e. $\Delta n_\mathrm{eff}(\lambda)-\Delta n_\mathrm{eff}(\lambda_0)$. In Fig.\,\ref{fig:sweep} the close match between our results and simulations considering dispersive models for the waveguide materials and the samples (dotted lines) can be observed. Absorption spectra for the three different scenarios were also recorded, but are not shown here due to insufficient accuracy. To provide reliable absorption spectroscopic measurements it would be beneficial to re-design the sensor addressing bandwidth and cross-talk, as well as to operate in wavelength regimes where samples with characteristic spectra can be detected.

\section{Conclusions} \label{sec:concl}

We have developed a simultaneous refraction and absorption sensor based on a modified MZI with a coherent phase retrieval working at a central wavelength of $1.55\,\mathrm{\upmu m}$. Our sensors have been fabricated in a commercial silicon nitride platform and evaluated by bulk sensing experiments with water-based samples at a fixed wavelength, showing state-of-the-art performance in the detection of both the real and imaginary parts of the complex refractive index. Preliminary spectroscopic experiments have been carried out, yielding promising results for dispersion measurements, which are specially suited for the analysis of liquid and highly concentrated samples \cite{Dabrowska2022}. We believe that our versatile architecture, which combines a precise and robust phase readout with useful absorption measurements, has a significant potential to be extended to MIR-compatible platforms, where many environmental and clinically relevant analytes have fingerprint spectra. Doing so will enlarge the retrieved information about the analyte, thus paving the way towards improved chemical sensors for the liquid and gas phases. 


\appendix
\section{Cross-talk model} \label{ap:fringes}

\begin{figure}[tb]
\centering 
\includegraphics[width=\linewidth]{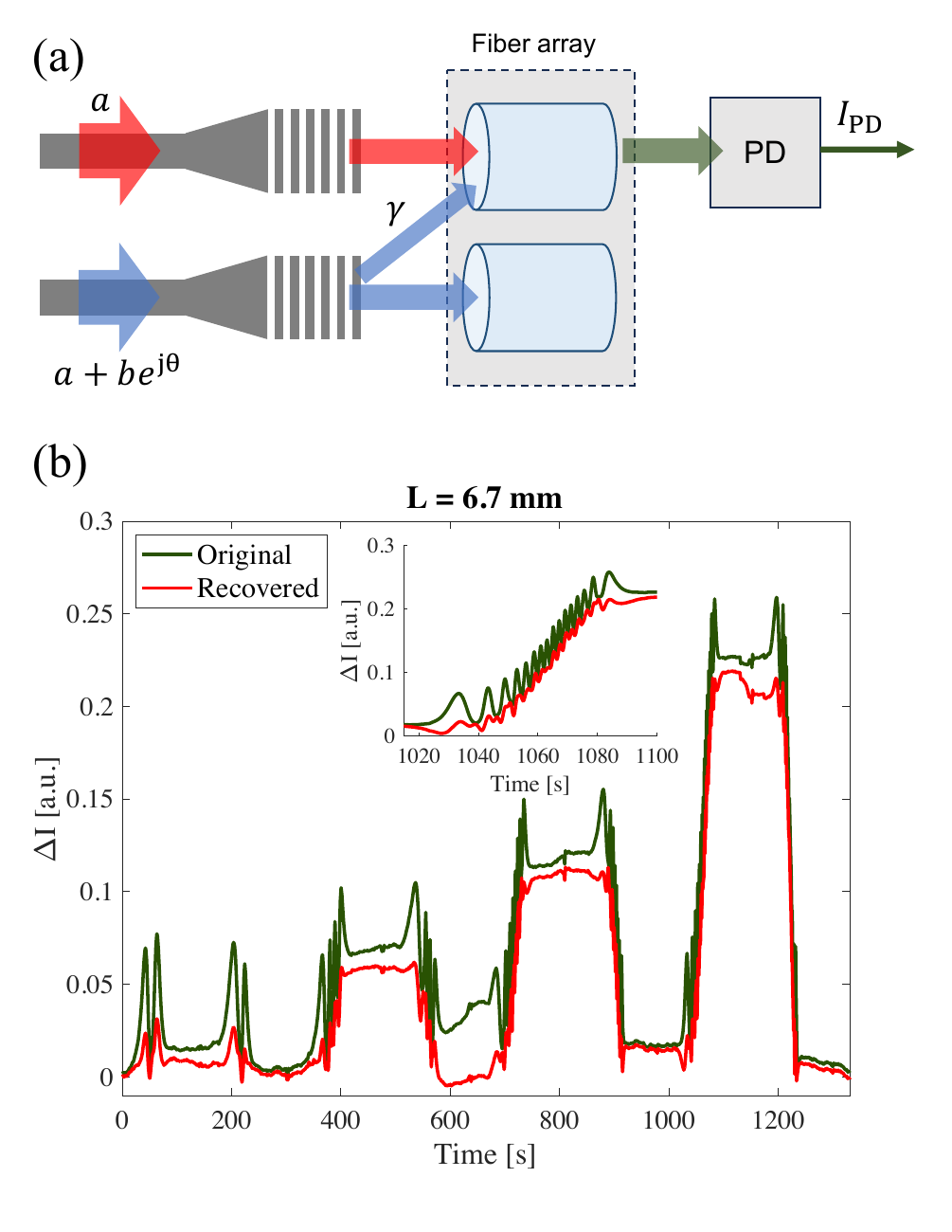}
\caption{(a) Schematic diagram of the cross-talk between adjacent outputs when out-coupled to the fiber array. (b) Intensity signal recovery for $\gamma=0.01$ and $\theta=45^\circ$. The visible reduction in the fringes supports our model.}
\label{fig:crosstalk}
\end{figure}

In section \ref{sec:exp} we specified that, in our setup, light coupling to/from the chip is performed by gratings coupled to an angle-polished fiber array.  There is a $127\,\mathrm{\upmu m}$ separation between adjacent fibers, and, consequently, neighbouring grating couplers are separated by the same distance. 
Here we model the cross-talk between adjacent outputs, which is particularly critical for the absorption measurements. Referring to Fig. \ref{fig:crosstalk}(a),
let $a$ be the electromagnetic field signal after travelling through the sensor arm, so that $I_\mathrm{s}=|a|^2$, and $b$ its counterpart for the reference arm. The upper phase output is a combination of $a$ and $b$, so that $I_1=|a+be^{\mathrm{j}\theta}|^2$, where $\theta$ accounts for the $120^\circ$ introduced by the $2\times3$-MMI plus any deterministic (i.e., not influenced by the sensing experiment) phase variations [see Fig. \ref{fig:diagram}]. If we consider a residual cross-talk coefficient $\gamma$ between the sensing intensity ($I_s$) and the upper phase outputs ($I_1$), as schematized in Fig.\,\ref{fig:crosstalk}(a), then, the photodetected intensity can be expressed as 
\begin{equation}
    I_\mathrm{PD} = \left| a + \gamma(a+be^{\mathrm{j}\theta})\right|^2.
    \label{eq:Ictalk_0}
\end{equation}
By expanding Eq.\,\eqref{eq:Ictalk_0} under the assumptions of $\gamma\ll1$ (weak coupling) and of $|b|\approx1$ (constant reference signal amplitude), we get the approximation
\begin{equation}
    I_\mathrm{PD} \approx (1+2\gamma)|a|^2+2\gamma\cos{(\Delta\phi+\theta)}|a|
    \label{eq:Ictalk_eq},
\end{equation}
where $\Delta \phi$ is the phase difference from Eq. (\ref{eq:DeltaPhi}), which we measure directly from the three phase outputs $I_{1,2,3}$. 
The desired sensing output intensity $I_\mathrm{s}$ can be thus recovered by solving the quadratic equation for $|a|$ and judiciously adjusting the parameters $\gamma$ and $\theta$. Figure \ref{fig:crosstalk}(b) shows a fragment of the original and recovered signals for $\gamma=0.01$ and $\theta=45^\circ$. The observed mitigation of fringes validates our modelling approach.


\section*{Acknowledgement}
This work has received funding from the Ministerio de Universidades, Ciencia e Innovación (FPU19/03330, PID2019-106747RB-I00), the Junta de Andalucía (Agencia Andaluza del Conocimiento PY18-793, Consejería de salud y familia PIN-0113-2020), and project TED2021-130400B-I00/ AEI/10.13039/501100011033/ Unión Europea NextGenerationEU/PRTR.


\end{document}